# Initial Performance of a Long Axial FOV PET with TOF and DOI capabilities: IMAS system

Antonio J. Gonzalez, *Senior Member, IEEE*, Alvaro Anreus-Valero, David Sanchez, Santiago Jiménez-Serrano, Marta Freire, Andrea Gonzalez-Montoro, *Member, IEEE*, Edwing Y. Ulin-Briseno, Neus Cucarella, John Barrio, Andrew Laing, Jorge Álamo, Julio Barbera, Luis F. Vidal, Marc Gil, Jose M. Benlloch, Alfonso Rios, Luis Marti Bonmati and Irene Torres-Espallardo

*Abstract* —This work summarizes the design, construction, initial performance evaluation and pilot clinical results of the IMAS system, a long axial field of view (FOV), also known as total-body (TB-), positron emission tomography (PET) prototype scanner. This PET enables for the first time in TB-PET imaging, simultaneously time-of-flight (TOF) and depth-of-interaction (DOI) capabilities. The IMAS detector block is based on LYSO semi-monolithic scintillators, with individual slab sizes of 3 mm × 25 mm × 20 mm each. Arrays of 1×8 slabs are coupled to 8×8 Silicon Photomultiplier arrays. A proprietary readout reduces the 64 signals to only 16 outputs, preserving both 3D photon impact positioning and timing accuracy. IMAS has a total of 30,720 channels. PETsys electronics is used for data acquisition. The IMAS geometry is based on 5 rings of 10 cm each, with a 5 cm gap between them. It defines an axial FOV of 71 cm with a bore aperture of 82 cm. We report in this work the pilot tests of the system performance and the first clinical results.

We found that the system spatial resolution remained below 4 mm across the entire FOV, even at the off-radial position of 30 cm. A coincidence time resolution with a small size $^{22}$Na source of 560 ps FWHM was measured. A sensitivity of 56.54 cps/kBq is in good agreement with previous simulation studies; however, the noise equivalent count rates performance (79 kcps at 3.26 kBq/mL) was significantly lower than expected, likely due to a data transfer bottleneck between the system and the acquisition workstation. Finally, a comparison of one of the imaged patients with a commercial TOF PET/CT scanner is also provided, pinpointing an improved tumor identification for IMAS, and the advantages of TOF and especially DOI capabilities.

*Index Terms*—DOI, Semi-monolithic scintillators, long axial FOV, TB-PET, TOF

## I. INTRODUCTION

TOTAL-BODY positron emission tomography (TB-PET) systems are very attractive given their unique feature of an increased acquisition field of view (FOV). Long-axial FOV (LAFOV) or TB-PET brings different advantages when compared to standard (conventional) whole-body (WB) PET systems, whose axial FOV (aFOV) is in the range of 25-30 cm [1]. One of these features is the significant increase in system sensitivity, given by the increase in angular detection coverage. Clinically, improving PET sensitivity enables a reduction of the patient dosage, scanning time, or a compromise between both [2]. A boost in sensitivity is also possible by using thicker detectors. However, this does not increase the aFOV and negatively impacts their performance in terms of energy and time resolution, also increasing the uncertainties in the determination of the photon impact depth of interaction (DOI), among others.

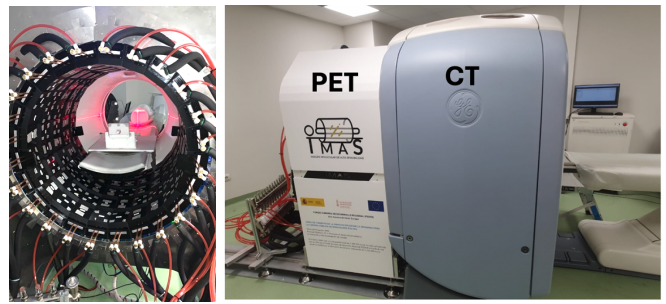

Fig. 1. Photographs of the IMAS TB-PET system and CT in tandem configuration.

One of the key advantages of TB-PET systems is the ability to simultaneously image multiple organs and to evaluate in a single bed their metabolic kinetics behavior or even the

This work was partly supported by the IMAS project launched by the Consellería de Sanitat Universal I Salut Publica of the Government of Valencia Region, announced in BOE 328, December 28, 2020, co-funded at 50% by the European Regional Development Fund (ERDF). E. U.-B. is supported by the Santiago Grisolía grant of the Generalitat Valenciana No. CIGRIS/2023/029. A. G.-M. was supported by the Spanish Ministry of Science and innovation and the European Social Fund Project: RYC2021 031744. M. F. is supported by the CIAPOS Program for Researchers in Postdoctoral Phase Grant of the Generalitat Valenciana under No. CIAPOS/2023/141. M. G. is also supported by the pre-doc grant ACIF of the Generalitat Valenciana No. CIACIF/2024/092. *(Corresponding author: Antonio J. Gonzalez).*

This work involved human subjects in its research. Approval of all ethical and experimental procedures and protocols was granted by the *Comité de Ética de la Investigación con Medicamentos CEIM – Hospital Universitario y Politécnico La Fe* under Protocol 17-07-2024, and performed in line with the actual laws and the Helsinki declaration of the world medical association.

A.J. Gonzalez, A. Anreus, S. Jimenez-Serrano, M. Freire, A. Gonzalez-Montoro, E. Ulin-Briseño, N. Cucarella, J. Barrio, A. Laing, L.F. Vidal, M. Gil and J.M. Benlloch, are with the Instituto de Instrumentación para Imagen Molecular (I3M), Centro Mixto CSIC—Universitat Politècnica de València, 46022 Valencia, Spain (e-mail: agonzalez@i3m.upv.es).

D. Sanchez, J. Alamo and J. Barbera are with Oncovision SA, 46022 Valencia, Spain.

A. Rios is with Full Body Insight, 46980 Paterna, Spain.

I. Torres-Espallardo and L. Marti-Bonmati are with Hospital Universitari i Politècnic La Fe, 46026 Valencia, Spain.

All authors declare that they have no known conflicts of interest in terms of competing financial interests or personal relationships that could have an influence or are relevant to the work reported in this paper.

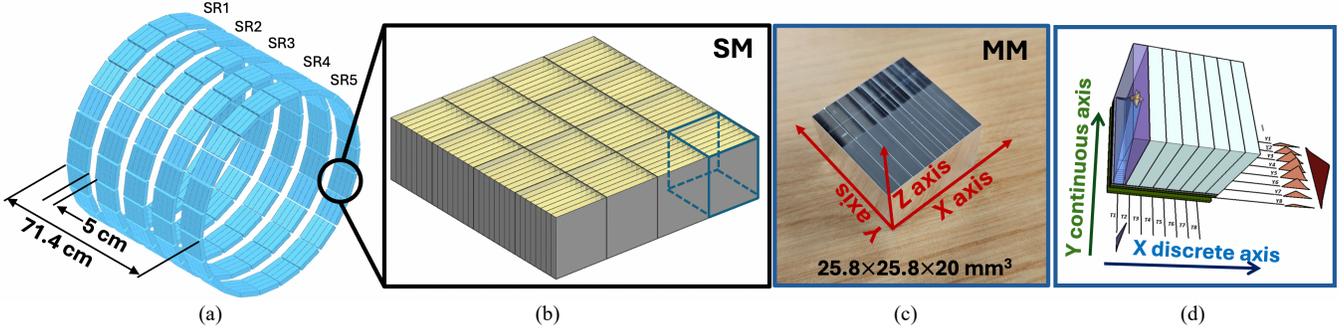

Fig. 2. System geometry, including modules and scintillator details. (a) Rings and gap dimensions. (b) Sketch of one super-module. (c) Photograph of the 1×8 array of slabs. (d) Sketch of the 64:16 reduction scheme of signals.

correlation between them within the same exploration [3]. There are many more advantages related to TB-PET as it has been detailed in other publications [4], such as imaging using short half-life isotopes like $^{11}C$ or $^{13}N$ [5], allowing one for longitudinal studies given the low administered dose [6], or simply improving the contrast by the gain in sensitivity.

There exist few TB-PET systems that are commercially available, namely the Vision Quadra from Siemens with 1.06 m aFOV [7], the Panorama from United Imaging [8] featuring 1.5 m aFOV and the former and precursor of this technology the uExplorer also commercialized by United Imaging with up to 2 m aFOV [9]. Another long-axial PET is the PennPET, currently reaching about 1.5 m [10], based on digital Silicon Photomultipliers but it is not commercially available. All these systems are based on pixelated crystal arrays without DOI capabilities, but with time of flight (TOF) information in the range of 180-400 ps. Another research-oriented LAFOV system is the TB-J-PET, axially arranging plastic scintillator strips read out at their ends by Silicon Photomultipliers (SiPMs), reporting TOF values of about 500 ps [11]. Novel capabilities, such as positronium imaging and quantum entanglement, have been tested in a 50 cm axial coverage and portable prototype [12]. Some of the performance of the existing TB-PET scanners will be summarized later in the Discussion section.

Regarding system performance, TOF capabilities further increase the effective sensitivity beyond the intrinsic physical sensitivity. Moreover, DOI information is of high relevance near the edges of the scanner FOV, where parallax errors become larger. Nevertheless, in contrast to pre-clinical imaging, where DOI is almost mandatory [13], its implementation in the clinical WB-PET environment has been less pursued, as patients are placed in the center of the scanner and the lesions of interest are typically larger. However, DOI is still important in both WB- and TB-PET since it can return a homogeneous spatial resolution response [14][15] and allows one to design systems with a smaller aperture (cost reduction) and a reduced performance deterioration at the edges of the FOV.

In this work, we present the design of a unique LAFOV clinical PET system with simultaneous DOI and TOF capabilities, named IMAS (meaning: High Sensitivity Molecular Imaging in Spanish). We provide insights about the system geometry, including the detector response, and pilot tests of the system performance, most of them inspired by the NEMA 2018 protocol [16]. Moreover, this work will present some of the first clinical results obtained with IMAS and compared them with those of a conventional WB-PET with TOF capabilities.

## II. MATERIALS AND METHODS

### A. System geometry and detector design

The IMAS PET system is built using a scintillator geometry based on semi-monolithic (slab) crystals [17][18][19][20]. The scanner defines a bore diameter of 82 cm and 71 cm axial coverage by assembling 5 so-called super-rings (SR) of 10 cm length each, with 5 cm gaps (see Fig. 2 (a) and [21] for further details). It is aligned to an GE Brightspeed 16 Computed Tomography (CT) working in tandem mode. The detector units are composed of LYSO mini-modules of 8 slabs of 3 mm × 25 mm × 20 mm each coupled to an array of 8×8 SiPMs of $3 \times 3$ mm$^2$ elements. We have arranged $4 \times 4$ of these -minimodules (MM) in so-called super-modules (SM). Each scanner ring is composed of 24 SM (see Fig. 2). The slabs are aligned with the axial axis of the scanner. Several surface crystal treatments and photosensor types were tested to find the best compromise between impact position accuracy (including DOI), energy performance, and timing [19][22]. The chosen configuration uses the photosensor model S13361-3050AE-08 (S13) from Hamamatsu Photonics and the treatment to the slab walls in which all faces were covered with Enhanced Specular Reflector (ESR).

### B. Electronics

The IMAS PET system includes two main electronic implementations:

(i) A novel and proprietary reduction readout is connected to each 8×8 SiPMs array of the MM. This multiplexing readout scheme reduces the number of output signals from 64 (number of SiPMs) to 8+8 (row and column channels), without almost impacting performance [23], see sketch in Fig. 2 (d). This is obtained by a circuitry that sums the information across the slabs, ensuring the integrity of the charge collection to retrieve accurate monolithic and DOI information, and additionally optimizes the sum of signals along the slabs faces to enhance the timing capabilities. The timing information is determined from these last signals. This concept has also successfully been used in two similar detector blocks for brain PET imaging [24][25]. Notice that we estimated that for most of the impacting annihilation photons we transmit 11 hits, 8 corresponding to the signals across the monolithic directions and (y-coordinates) 3 additional ones parallel to these faces (x-coordinates). That means a total of 22 signals per coincidence event.

(ii) PETsys technology (Lisbon, Portugal) is used for the front-end and data acquisition electronics. The multiplexed signals are amplified and shaped before being fed to the TOFPET2 ASIC [26]. The FEM256, a board allocating four ASICs (64 channels each), is a key component in this architecture since the signals of each SM (4×4 MM ×16 reduced signals = 256) are read by one FEM256. The digitized energy and timestamp data for each gamma event are serialized and sent to a FEB/D 4096. A dedicated Clock/Trigger board is responsible for distributing the master clock to eight FEB/D units. This board also validates trigger coincidences, using a look-up table (LUT) to define the allowed coincident regions. Finally, the data from all FEB/D boards are transmitted to the PETsys DAQ board connected to a PC via PCI-E, where the raw data is written to a fast NVMe disk.

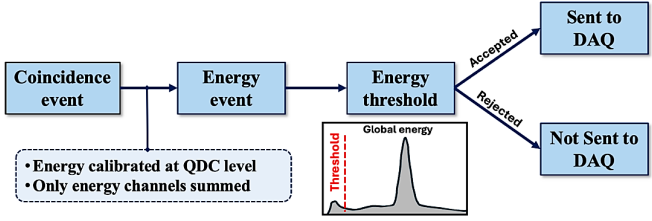

Fig. 3. Diagram of the global energy threshold implemented at the FPGA of the FEB/D.

During the IMAS development phase, two improvements from PETsys were introduced, namely the use of a second DAQ board and an online energy filter, both aimed towards a larger count rate performance. The second DAQ effectively increases the data-handling capacity of the system. Following this hardware implementation, the event processing pipeline was optimized at the firmware level. To effectively manage the high rate of signals generated by the detectors and to increase the true event throughput to the DAQ, a global energy threshold was implemented directly at the FPGA of the FEB/D (see Fig. 3). This feature processes the summed energy from the corresponding detector channels (after QDC-level calibration) and applies a pre-defined threshold based on the signal levels [27]. Events falling below this threshold —typically low-energy noise— are rejected at the hardware level before transmission. This minimizes the amount of unnecessary data sent to the PC/DAQ, reducing data-link saturation while allowing for a higher rate of relevant events.

### C. System calibration

The x-(pixelated) coordinates of the gamma ray impact are directly inferred from the triggered slab, while the y-(monolithic) and DOI-(z) positions require specific algorithms to mitigate edge effects. To address this, we implemented a Neural Network (NN) methodology based on two groups of Multilayer Perceptrons (MLPs): one dedicated to predicting the y-(monolithic) impact position (MLP$_Y$) and another focused on estimating the DOI (MLP$_{DOI}$). The MLPs have 8 inputs that correspond to the 8 signals along the monolithic direction after equalization and one output, the predicted y-(monolithic) or DOI positions (see Fig. 4).

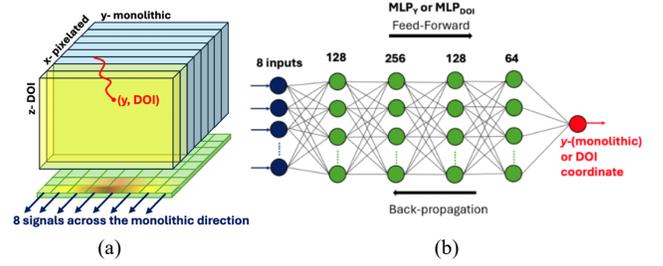

Fig. 4. (a) Sketch of a MM showing the signal projections used as MLP input. (b) Sketch of the MLP structure.

Ideally, experimental data of every SM of the scanner would be acquired to obtain the ground truth information to train the MLPs. However, this process is time-consuming and computationally costly, and, thus, experimental data were acquired only from a reference SM (see Fig. 5). A slit collimator with a collimation aperture of 500 µm, and two $^{22}$Na sources (0.25 mm active diameter and about 5 µCi each) were placed between two identical SM. A translational stage motor was used to move the reference SM. For the y-(monolithic) coordinate, the SM was moved along the monolithic direction in 1 mm steps. For DOI training, the SM was rotated 90° to place the source along the lateral faces of the slabs and also moved in 1 mm steps. A separate dataset was generated for each MM and for each direction. The dataset was then randomized and divided into training, evaluation, and test sets (70%, 10% and 20%, respectively). Data from MM0 was used for training and evaluation to avoid overfitting, while data from the remaining MMs was reserved for testing the MLP performance.

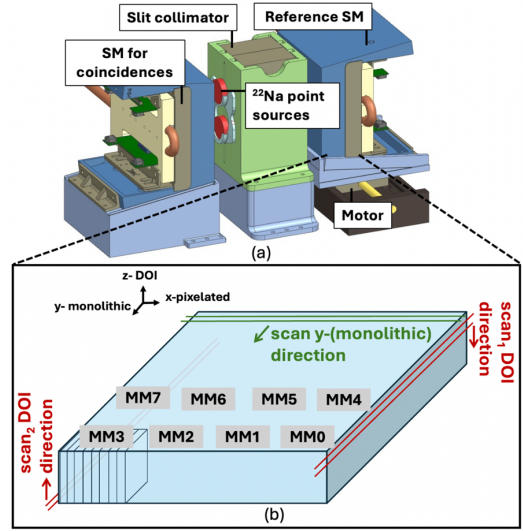

Fig. 5. (a) Experimental set-up used for the NN training. (b) Sketch of the different scans acquired for the reference SM. Green lines show the scan along the monolithic direction, whereas the red lines depict the scans of the 4 MM at each side of the SM to train and evaluate the DOI.

We found that in certain MMs of the IMAS system, some of the eight signals across the monolithic direction exhibit issues (i.e., zero energy, different gain, etc.). To account for this variability, the data from MM0 was synthetically modified to generate multiple datasets reflecting these conditions, obtaining 12 classes in total, namely, a non-modified (original) and 11 modified datasets for each direction ($y$-(monolithic) and DOI). Therefore, a total of 24 different MLPs (referred to as categories) were trained. Notice that these MLPs were used to predict the $y$-(monolithic) and DOI impact positions in all the 15,360 slabs of the IMAS system by assigning each slab to an MLP category.

Regarding the robustness evaluation of the process, the data from the MMs of the reference SM (test dataset) were predicted using the assigned MLP category. The prediction accuracy was evaluated by calculating the Bias, Mean Absolute Error (MAE), and Full Width at Half Maximum (FWHM) of each slit position for each MM of the reference SM.

The spatial capabilities of the IMAS detectors were also evaluated by placing an FDG bar (700 mm long × 10 mm diameter) with low activity at the center of the system to obtain a uniform measurement. The $y$-(monolithic) and DOI impact positions for all the SMs were predicted as explained. Flood maps (2D histograms of the $y$ and DOI predicted coordinates) were generated for each MM of the system. From these flood maps, the profiles across the $y$-(monolithic) and DOI directions were obtained. To provide a quantitative metric of the spatial calibration quality for the entire system, the minimum and maximum limits of those predicted profiles were determined by identifying the first and last bins and applying a threshold based on the number of counts.

Since achieving accurate TOF performance requires correcting the intrinsic timing offset (skew correction) for each detector element to account for variations in electronic response and light collection, we employed an iterative timing calibration algorithm [28] to determine the timing offsets. The core principle assumes that the overall TOF-difference bias for any pair of coinciding detector elements (or slabs in this system) is the difference between the absolute time biases of the individual slabs. It requires measuring the offset of each individual slab against many opposing slabs, rather than measuring every possible slab pair. The iterative correction converges when all individual timing histograms are centered at zero. The result is a timing offset correction, stored as a LUT. For this process, the same bar was filled with 72.15 MBq of FDG placed at 160 mm off-radial position. The bar was smoothly rotated at a revolution time of 120 seconds. The obtained data was energy filtered (408 – 613 keV), and only lines of response (LORs) with a maximum aperture of 15° were allowed. Furthermore, to confirm this procedure and to determine the coincidence time resolution (CTR) of the system, an encapsulated $^{22}$Na source (164.7 kBq activity and 1 mm active diameter) was placed close to the center of the system and scanned for 2000 seconds.

### D. Image reconstruction

The standard image reconstruction protocol that has been implemented for IMAS, uses an accelerated GPU-based List-Mode Ordered Subset Expectation Maximization (OSEM) implementation, typically with 5 iterations and 5 subsets. Each 3D image consists of 391 × 391 × 346 cubic voxels of 2.06 mm. Each coincidence event is DOI corrected before generation of the List-Mode (LM) file by reassigning the measured LOR to the corresponding virtual detector pixels at its front face [29]. Moreover, we exploited the accurate timestamp differences (Δt) of coincidence events and discretized those into 7 TOF bins [30][31]. For each bin, separate coincidence and scatter histograms were generated, using 750 ps FWHM CTR as the effective temporal resolution in the reconstruction.

Attenuation correction was consistently applied using the corresponding CT acquisitions. Scatter correction using TOF followed the scaling method described in [32]. The two energy windows were defined as 358 – 409 keV for scatter estimation and 410 – 613 keV for true prompt coincidences. Random coincidence correction was not applied, as singles data is not currently stored. Data were also corrected for normalization, which is key in all systems but plays a more important role in scanners with gaps between detectors, see for instance reference [33]. For this process, the described fillable FDG bar was repeatedly placed inside a thin carbon fiber tube (to avoid any bending) at a radial off-center distance of 350 mm. It was filled with initial FDG activities in the range of 50 MBq, and data was acquired for about 5-6 hours each time. The mean number of counts acquired per LOR was 19.4 ± 17.4 for a total number of $1.52 \times 10^9$ LORs.

### E. Performance tests

The characterization of the IMAS system has been inspired by the NEMA 2018 protocol [16]. We have studied spatial resolution, sensitivity, count rate capabilities, and image quality, as well as TOF performance. The details of the acquisitions and methodology are explained in the following:

i) Spatial resolution. A $^{22}$Na source with 230 kBq activity and 1 mm active diameter was placed at the center and at 3/8 of the aFOV, and then data was acquired at 1, 10, 20 and 30 cm off radial positions, for 600 s each. The data were reconstructed using the Maximum Likelihood Expectation Maximization (MLEM) algorithm with 15 iterations. We report the FWHM, as well as the Full Width at one Tenth of the Maximum (FWTM) for all positions and directions.

ii) Sensitivity. A PMMA tube, 700 mm long and 3 mm inner diameter, was filled with 13 MBq of FDG and placed centered along the axial axis of the scanner. Then, 4 aluminum layers of 1.25 mm wall thickness each were sequentially added, starting with an inner diameter of 3.9 mm. Each measurement lasted 180 seconds. For each sleeve, total count rates were measured and corrected for radioactive decay. These corrected values were fitted to an exponential attenuation model to estimate the unattenuated count rate, as described in [16].

iii) Count rate capabilities. The count rate performance, including the Noise Equivalent Count Rate (NECR), was evaluated using a large high-density polyethylene solid phantom of 203 mm in diameter and 700 mm axial length. A 64 mm hole was drilled 45 mm off-center, into which a plastic tube filled with an initial activity concentration of 10.70 kBq/mL was inserted. Then, a total of 70 acquisitions of 120 seconds each were performed, with 900 seconds intervals between scans. True events were identified from the sinogram peak, while random and scatter contributions were estimated from the surrounding plateau region, following the approach also described in [16]. The energy resolution of the system is also provided at the NECR peak.

iv) <u>Energy and Coincidence Time Resolution.</u> The CTR as a function of the activity concentration has also been estimated using the data obtained with the NECR phantom [16]. The reported energy resolution was calculated using the same phantom at the activity concentration of the NECR peak.

v) <u>Image quality.</u> The torso phantom [16] was employed for these tests. See details with dimensions of the phantom in Fig. 6 (a). An activity ratio between hot and background regions of 1:8.3 was set, with a total activity of 37.42 MBq. The acquisition lasted 2000 seconds. The concentration of the spheres was 31.23 kBq/mL, whereas for the background concentration was 3.76 kBq/mL. The Contrast Recovery (CR), Background Variability (BV), and the lung insert accuracy corrections (AC) were obtained as described by the NEMA 2018 protocol for different numbers of iterations.

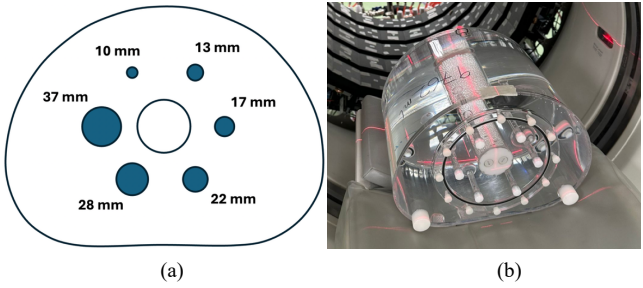

Fig. 6. (a) Sketch of the torso phantom as described by the NEMA 2018 protocol. (b) Photograph of the phantom on the IMAS bed.

### F. Clinical evaluation

This work also includes preliminary clinical data as a demonstration of the system diagnostics capabilities. The clinical validation of IMAS is currently undergoing with patients who are programmed for conventional PET/CT, at the hospital La Fe in Valencia (Spain). We are comparing our results with those obtained with the Philips Gemini TF-64 PET/CT, which uses an OSEM list-mode algorithm including TOF (585 ps FWHM) and enables a 3D volume of $144 \times 144 \times 130$ cubic voxels, each with 4 mm isotropic dimensions [34]. In particular, for this comparative purpose, we have selected a case of a female patient of 45 years and 51 kg with multiple and distant lesions. The patient was first scanned in the Philips PET/CT (1200 seconds acquisition) and about 90 minutes later in the IMAS, for also 1200 seconds and with an actual activity of 119.9 MBq.

Although IMAS includes the GE CT in a tandem configuration, see Fig. 1., to avoid any extra dose delivered to the first set of patients, we have used the CT data from the Philips PET/CT for attenuation correction. To co-register our PET with the Philips CT images, we employed a multi-stage registration approach using the Elastix framework [35]. The registration process consisted of two sequential stages: an initial translation transformation using geometric center initialization, followed by an affine transformation optimized through adaptive stochastic gradient descent with Advanced Mattes Mutual Information as the similarity metric. The registration was performed in a multi-resolution framework using grid-based sampling with B-spline interpolation for final resampling.

### III. RESULTS

### A. Detector performance

The selected detector configuration (S13 series from Hamamatsu and ESR covering all crystal walls of slabs) reports an average Detector Time Resolution (DTR) of 196 ± 13 ps (equivalent to 277 ps CTR), with an energy performance of 11 ± 1 % [19]. Regarding impact accuracy, although a slightly better performance was achieved when black paint was added to the lateral walls of the slabs, the performance when ESR is used was 2.8 ± 1.0 mm and 3.9 ± 0.3 mm, for the monolithic and DOI directions, respectively [22].

### B. System calibration

As an example, Fig. 7 (a) and (b) show the Bias, MAE, and FWHM for each slit position along the $y$-(monolithic) direction for MM4 and MM12 of the reference SM, respectively. The average and standard deviations are shown as inlets in these figures. Each of these MMs has been predicted using its corresponding category. MM12 is a detector block for which one of the 8 signals across the monolithic direction fails. The performance along the DOI direction for MM4 and MM12 is represented in Fig. 7 (c) and (d), respectively. Notice that the position at 0 mm is the entrance face, whereas 20 mm is the position closer to the photosensor.

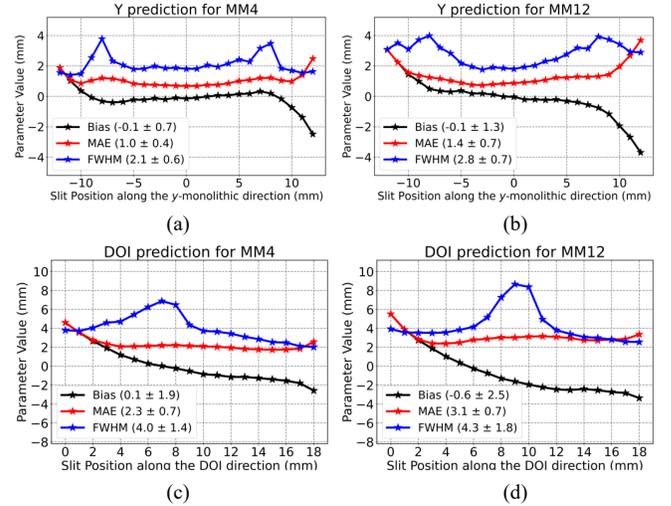

Fig. 7. (a) Bias, MAE, and FWHM along the $y$-(monolithic) direction for MM4. (b) Bias, MAE, and FWHM MM12 along the $y$-(monolithic) direction. (c) Bias, MAE, and FWHM along the DOI direction for MM4. (d) Bias, MAE, and FWHM along the DOI direction for MM12. The mean value over all slit positions is shown in the legend.

Fig. 8 (a) and (b) show the mean and standard deviation values (as error bars) of the MAE and FWHM considering all the slit positions for each MM from the reference SM in both the $y$-monolithic (24 positions) and DOI directions (19 positions), respectively. Here, each MM was predicted with its assigned category.

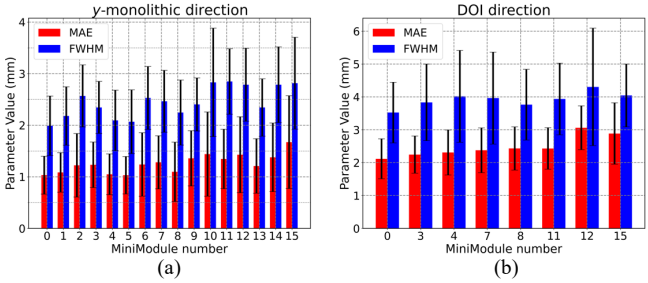

Fig. 8. Mean MAE and FWHM values for each MM from the reference SM for (a) the y-(monolithic) and (b) DOI directions.

Regarding the calibration performance of the whole system, a coincidence map for one SM and one MM is presented in Fig. 9 (a) and (b), showing the profiles along the y-(monolithic) and DOI directions. The bumps at the edges of the y-profile agree with the bias observed at the detector level, see Fig. 7 (a) and (b).

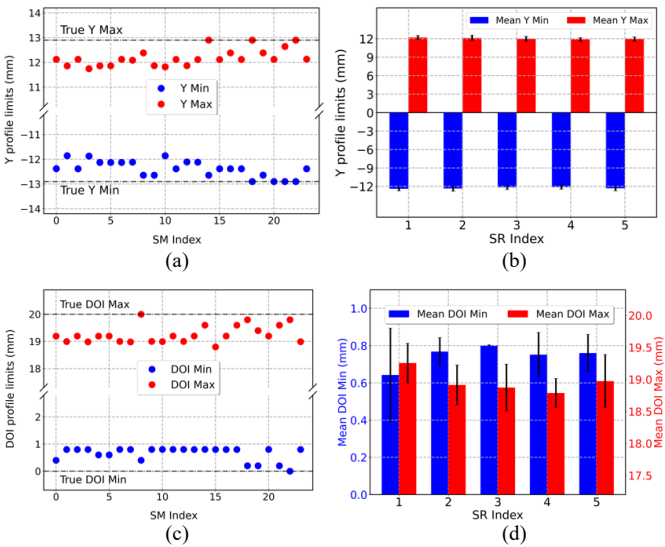

Fig. 9. (a) Flood map of the x-y predictions for one SM. (b) Flood map of the x-y predictions for one MM and profiles along the y-(monolithic) and DOI directions.

Fig. 10 (a) shows the minimum and maximum limits found for the profiles along the y-(monolithic) directions for each SM of the SR1. In Fig. 10 (b), the mean and standard deviation values are shown for each SR. The minimum and maximum limits obtained for the DOI profiles of each SM of SR1 are depicted in Fig. 10 (c) and (d), together with the mean and standard deviations of these limits, for each ring.

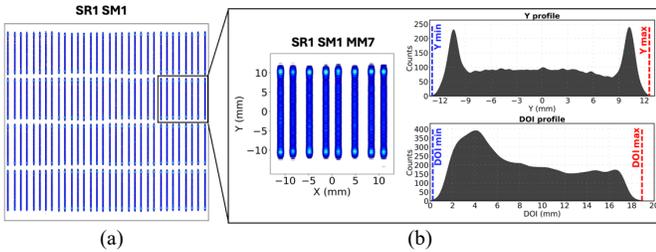

Fig. 10. (a) Min and Max limits of the y-profile for each SM of SR1. Notice that the value zero is the center of the slab in the monolithic direction. (b) Mean and standard deviation values of the y-profile for each SR of the IMAS system. (c) Min and Max limits of the DOI profile for each SM of the SR1. (d) Mean and standard deviation values of the DOI profiles for each SR.

Regarding timing performance, when estimating the skew correction using the data obtained with the rotational bar, all channels were calibrated after the 15-th iteration. The convergence criteria were set to stop when the centroid error felt below 10 ps or the change between successive iterations was less than 2 ps. The selection of the relaxation factor λ was tuned to optimize convergence, with values ≥ 0.6 being most effective. As seen in Fig. 11 (a), the centroid error values for all channels quickly converge towards zero as the number of iterations increases. This calibration process significantly improved the system performance, reducing, for instance, the CTR FWHM for a $^{22}$Na point source from $2863 \pm 35$ ps (uncorrected) to $565 \pm 5$ ps, see Fig. 11 (b).

## C. System performance

### a. Spatial resolution

Fig. 12 (a) shows the MLEM reconstructed images of the $^{22}$Na source at the axial center of the FOV and at different radial positions, without and with DOI correction. Fig. 12 (b) shows the spatial resolution (FWHM) values of the $^{22}$Na source along the radial directions, highlighting the effect of misposition when DOI corrections are not implemented. Quantitatively, there is a 6 mm shift towards the scanner center at 30 cm off-radial position when DOI is not corrected, found as the difference between the vertical lines at this position.

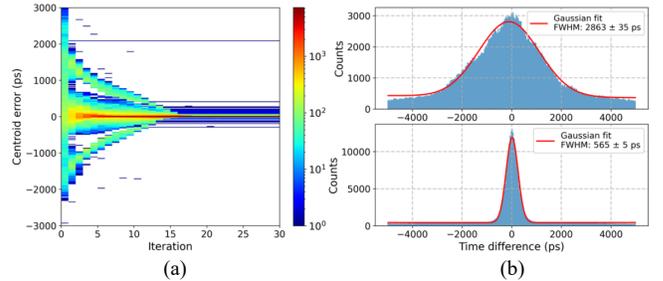

Fig. 11. (a) Centroid error of each channel versus the number of iterations with a relaxation factor of 0.9. (b) Timing spectra without and with skew correction for a $^{22}$Na source at the center FOV.

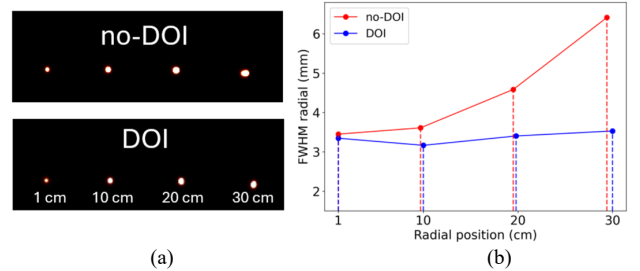

Fig. 12. (a) Reconstructed images of the $^{22}$Na source and (b) radial FWHM spatial resolution values along the radial direction at the center of the aFOV and at different radial positions, without (top) and with (bottom) DOI correction.

Table I reports the measured spatial resolution for all the different positions and directions, using MLEM and including DOI correction. IMAS has shown a homogeneous spatial resolution across the entire FOV with $3.3 \pm 0.5$ mm in the radial direction at 1 cm transaxial offset, and just degrading by about 0.2 mm at 30 cm. Notice that all available clinical systems, both WB- or TB-PET, only provide these values at a maximum offset of 20 cm. The average spatial resolution (axial center) for the three space components and positions is also $3.3 \pm 0.5$ mm (not including the values at 30 cm offset) The values reported at 3/8 of the axial FOV are also very homogeneous, resulting in $3.2 \pm 0.4$ mm on average.

TABLE I. SPATIAL RESOLUTION. MLEM 15 ITERATIONS.

| Position | Axial CFOV | | | | Axial 3/8 FOV | | | |
|---|---|---|---|---|---|---|---|---|
| | 1 cm | 10 cm | 20 cm | 30 cm | 1 cm | 10 cm | 20 cm | 30 cm |
| | FWHM (mm) | | | | | | | |
| **Radial** | 3.34 | 3.17 | 3.41 | 3.53 | 3.21 | 3.12 | 3.15 | 3.31 |
| **Tang** | 3.77 | 3.67 | 3.94 | 4.05 | 3.62 | 3.59 | 3.72 | 3.99 |
| **Axial** | 2.67 | 2.64 | 2.79 | 2.76 | 2.74 | 2.64 | 2.59 | 2.70 |
| | FWTM (mm) | | | | | | | |
| **Radial** | 6.10 | 5.77 | 6.20 | 6.43 | 5.84 | 5.68 | 5.73 | 6.02 |
| **Tang** | 6.88 | 6.69 | 7.18 | 7.40 | 6.60 | 6.54 | 6.77 | 7.27 |
| **Axial** | 4.87 | 4.82 | 5.10 | 5.03 | 5.00 | 4.81 | 4.71 | 4.93 |

### b. Sensitivity and count rates

The sensitivity profile is depicted in Fig. 13 (b), together with a photograph of the experimental set-up, resulting in 56.54 cps/kBq, very similar to the predicted value using simulation data, as described in [21]. Moreover, placing a source at the center of the scanner, we measured a peak sensitivity of 7.6%. That is about 4-5 times larger than existing conventional PET/CT scanners with 25-30 cm aFOV coverage.

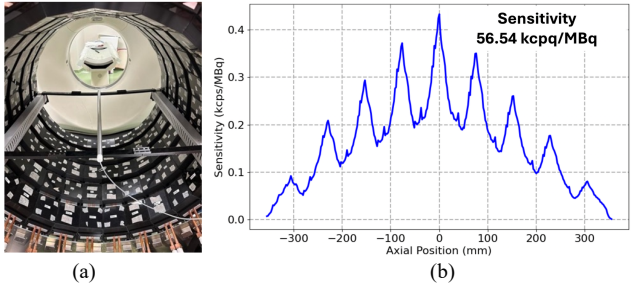

Fig. 13. (a) Photograph of the set-up with the filled tube and aluminum sleeves centered in the transaxial FOV and (b) measured sensitivity profile.

Fig. 14 shows the count rate performance for the IMAS system. The peak NECR was found at 79 kcps for an activity concentration of 3.26 kBq/mL. We observe a rather linear trend up to about 3 kBq/mL.

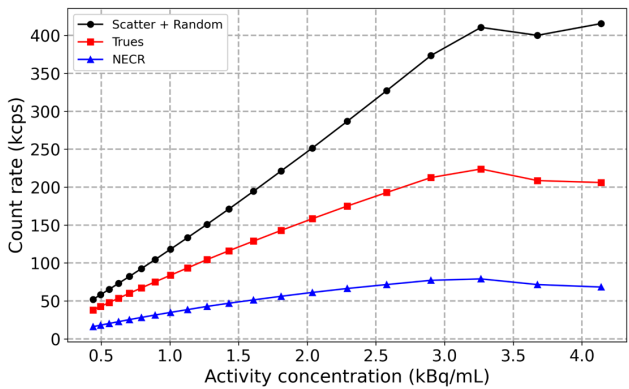

Fig. 14. Count rate performance as a function of the activity concentration. Trues, NECR, and Scatter + Random events are depicted using red squares, blue triangles, and black circles, respectively.

### c. CTR resolution

Fig. 15 (a) presents the CTR values of the system as a function of the activity concentration, as described in the NEMA 2018 protocol. We observe a degradation of the CTR with the activity concentration, varying from about 550 ps to 725 ps in the range of 0.4 to 4.2 kBq/mL. A CTR of 690 ps FWHM was obtained at the activity concentration of the NECR peak of 3.26 kBq/mL. Fig. 15 (b) depicts the energy profile at such NECR peak, resulting in an energy resolution of 12.8%.

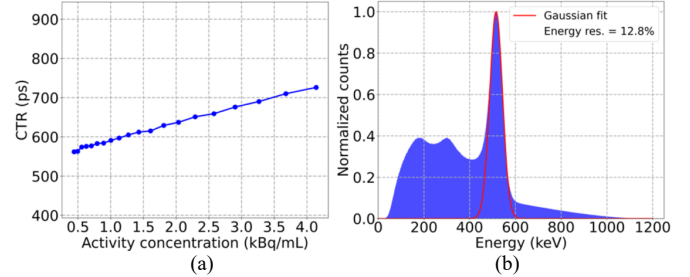

Fig. 15. (a) CTR values as a function of the activity concentration using the NECR phantom. (b) Energy profile and fit at the NECR peak.

### d. Image quality

Fig. 16 (a) shows the three views of 1 slice (2.06 mm) of the measured Torso phantom, together with the plot for the CR vs BV, for all spheres in the phantom and for reconstruction iterations (OSEM) ranging from 3 to 15. They included normalization, as well as attenuation and scatter corrections, but not randoms. In Fig. 16 (b) we observe a smooth CR improvement for all spheres when increasing the number of iterations, at the cost of some worsening of the BV. Values near 80% are found for the largest sphere of 37 mm, whereas the smallest one of 10 mm varies from around 30 to 40%. Iteration 5 was selected after a study in which a global metric is generated using a size-weighted average, balancing sphere size and recovery-background coefficient.

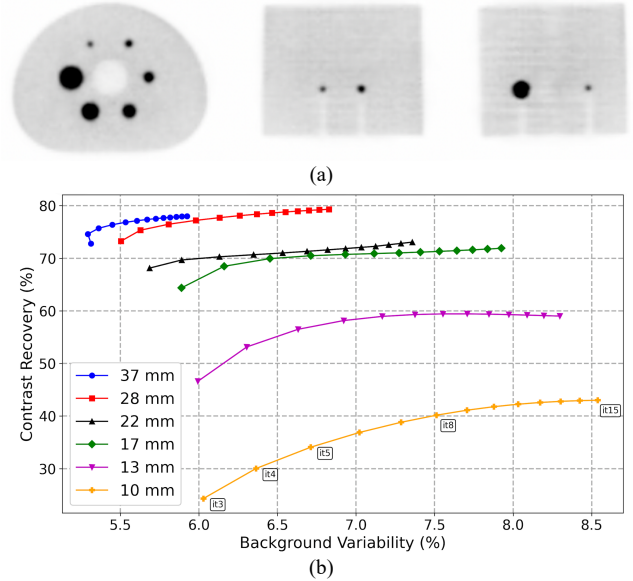

Fig. 16. (a) Views of the reconstructed images of the torso phantom (OSEM 5 iterations, DOI and TOF included) (b) Contrast recovery versus background variability, for all spheres in this phantom, as a function of the number of iterations.

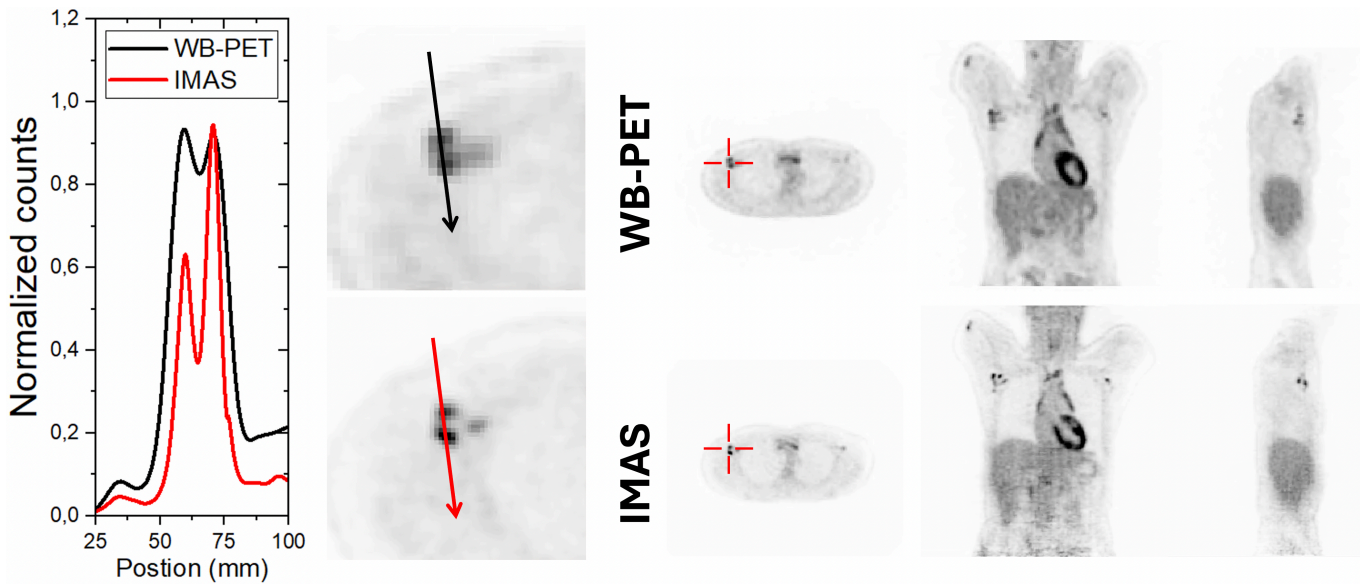

Fig. 17. Top row shows the reconstruction using the default parameters of the Philips Gemini-TF PET, whereas the bottom row shows the images obtained with the IMAS (OSEM 5 iterations, DOI and TOF included). In the three views, the IMAS images have not filter applied. The profiles (left) across the lesions of the zoomed images are normalized to the maximum. The co-registration of the two image sets was done manually.

### D. Clinical Results

Fig. 17 shows the reconstructed images of the patient obtained with the IMAS and Philips Gemini TF-64 systems. The preliminary image quality of the IMAS seems to be superior to the one provided by the conventional PET/CT, in terms of identification and definition of structures. In this patient, a multi-centric lesion was observed in the axilla, which was clearly resolved with IMAS, but blurred with the conventional PET/CT. The profiles across two lesions are also depicted in Fig. 17. Despite possible misalignments in the co-registration of the images, the identification of the two lesions is more evident for IMAS, with an increased signal-to-noise ratio (SNR).

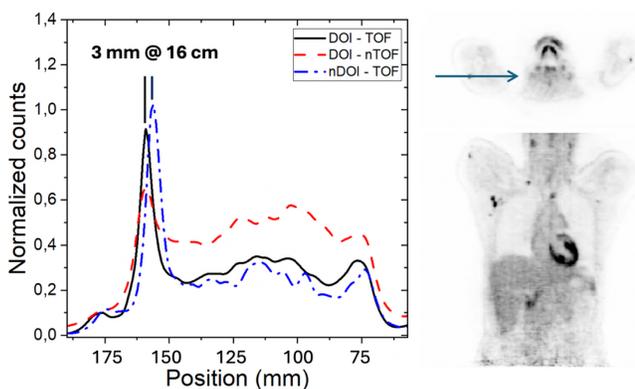

Fig. 18. Profiles along a lesion at 16 cm from the center of the scanner, for three IMAS reconstruction cases as depicted in the inlet legend. nTOF and nDOI stand for no TOF and no DOI considered during the reconstruction processes.

To highlight the advantages of the DOI capabilities, we have selected a lesion of the same patient that was about 16 cm away from the scanner center. For this lesion, we can observe that when DOI is not activated, a misposition of as much as 3 mm is found. The importance of the DOI is also observed in Fig. 12 where for source position that is 30 cm away for the scanner center is wrongly localized by 6 mm.

### IV. DISCUSSION

System calibration: IMAS is a first-generation prototype where the design and implementation processes were highly contributed by our research institute (i3M), in collaboration with three local companies (Full Body Insight, Oncovision, and Quibim). We are aware of some missing channels, most likely due to poor or broken connections between the boards. We might recover them by disassembling detector parts, but this will cause a long stop in the usage of the scanner, thus both researchers and clinicians prefer not to carry out until more evidence of the advantages of the technology is proven. Note, however, that these only represent 3.9% of all signals.

Spatial resolution: The measured average spatial resolution at the center of the axial FOV for all components of $3.27 \pm 0.48$ mm compares to $3.26 \pm 0.56$ mm, $3.96 \pm 0.81$ mm and $4.06 \pm 0.74$ mm, for the uMI Panorama, Siemens Vision Quadra, and PennPET, respectively [7][8][10]. Notice that the PennPET and IMAS data is obtained with an iterative reconstruction algorithm, whereas the Quadra and Panaroma used analytical ones. Considering the radial direction only, the IMAS performance is about $1.5 - 2$-fold better at 20 cm off-radial position, as observed in Fig. 19. Moreover, with the published data we estimated their possible performance at 30 cm using a parabolic fit. When these results are compared to the IMAS, they show a factor $2.5 - 3$ worst performances at 30 cm off-radial position. That is, other systems seem to degrade to $7 - 9$ mm FWHM. Table II summarizes the performance of IMAS and compares this with existing LAFOV clinical systems based on inorganic scintillators, including the 2 m long uExplorer.

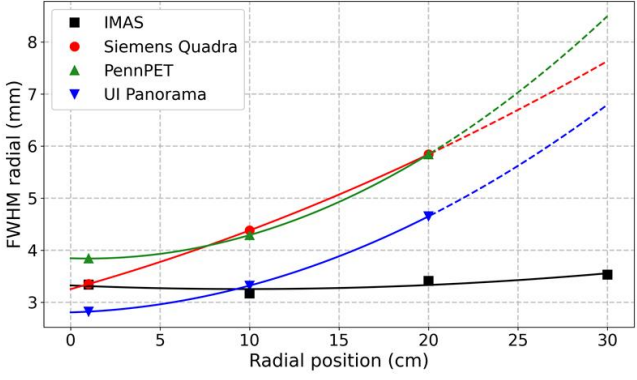

Fig. 19. Spatial resolution comparison of IMAS with other LAFOV scanners. The dotted lines are extrapolated using a second-order polynomial using the published data.

**Time resolution performance:** As observed in Fig. 11 the CTR of a low activity source was $564 \pm 5$ ps, but looking to the CTR versus activity curve (Fig. 15) we found 690 ps at the NECR peak of 3.26 kB/mL. An improvement to the range of 350 ps – 400 ps should be reachable after a more accurate time skew calibration, since the individual detectors reached such performance. Notice that this work implements the use of semi-monolithic crystals and a highly multiplexing circuitry, instead of pixelated, challenging the timing estimation.

TABLE II. COMPARATIVE DATA. RR STANDS FOR RADIAL SPATIAL RESOLUTION AT AXIAL CENTER, S FOR SENSITIVITY WITH 70 CM PHANTOM, CRYSTAL REFERS TO SCINTILLATOR HEIGHT, CTR IS GIVEN AT NECR PEAK, *TOTAL FOV, **SEMI-MONOLITHIC, ***ESTIMATED FROM PLOT IN REFERENCE [8]. CR AND BV FOR uEXPLORER DERIVED FROM PLOT IN [35].

| Reference | This work | Quadra [7] | PPET [10] | uMI [8] | uExplorer [35] |
|---|---|---|---|---|---|
| Axial (cm) | 71 | 106 | 142 | 148.2 | 194 |
| Bore (cm) | 82 | 82 | 76.4 | 76* | 78.6 |
| Crystal (mm) | 20** | 20 | 19 | 18.1 | 18.1 |
| DOI | YES | NO | NO | NO | NO |
| RR 1cm (mm) | 3.34 | 3.35 | 3.9 | 2.82 | 3 |
| RR 20cm (mm) | 3.41 | 5.84 | 5.6 | 4.65 | 4.7 |
| S (cps/kBq) | 56.54 | 82.6 | 140.2 | 176.3 | 174 |
| NECR (kcps, kBq/cc) | 79 @ 3.26 | 1613 @ 27.49 | 1550 @ 25 | 3350 @ 57.57 | 1524 @ 17.3 |
| CTR (ps) | 690 | 228 | 256 | 195*** | 430 |
| CR range (%) | 34-76 | 78-93 | 37-75 | 68-94 | 50-100 |
| BV range (%) | 5-7 | 1-2 | 2-3 | 1-2 | 1-4 |

**Rates:** The determined NECR of 79 kcps at 3.26 kBq/mL is significantly lower than other LAFOV systems, even if compared to WB-PET scanners. As reported, this does not seem to be a limitation to image patients in conventional conditions with $^{18}$F-FDG, but we are aware that for dynamic or high activities (as used for $^{11}$C studies, for instance), this might be a limitation. To mitigate this and maximize peak and data rates, our roadmap includes transitioning to a distributed acquisition framework. By implementing a dedicated workstation per individual FEB/D (totaling 5 to 8 units), we will parallelize disk-write operations, which is apparently the main bottleneck of the system.

**Image quality:** The calculated CR values were somehow lower than expected even for the largest spheres. Plotting CR values as a function of the BV for the different spheres and reconstruction iterations proved to be a convenient way to determine the appropriate number of iterations. The improvement of the CR and worsening of BV with the number of iterations was expected. We envisage a boost in the CR values to be in the range of 80-90% by improving the normalization procedure, highly impacting systems enabling ring gaps. For instance, we are working on the implementation of a component-based approach, as well as by including random corrections that might impact the quality of the images, given the huge amount of lutetium present in the scanner.

## V. CONCLUSION

In this work, we summarize the performance parameters of the IMAS system, a TB-PET with an axial coverage of 71 cm that simultaneously provides TOF and DOI capabilities. These features are achieved using an alternative scintillator geometry to the crystal arrays, that are the semi-monolithic. IMAS is a system that saves part of the cost, when compared to other existing TB-PET, by using an architecture that includes gaps, and a reduction readout scheme without almost compromising the detector performance.

Accounting for DOI information results in a homogeneous spatial resolution better than 4 mm in the whole FOV, extending to 30 cm off-radial position.

We are aware of the current limitations regarding count rates, due to the electronics chain, but this has not impacted our current results. This work was a proof-of-concept system for which there were detector constrains and data bandwidth not fully known in the design phase. We are nevertheless working on solutions to improve such performance.

Clinically, the increased sensitivity and homogeneous spatial resolution response in our preliminary acquisitions significantly improve the diagnostic prediction when compared to conventional PET. The presented patient data showed both multicentric structures near the right axilla, and additional lesions located at up to 16 cm, highlighting the ability of IMAS to resolve these lesions with high accuracy and precision.

## ACKNOWLEDGMENT

Authors thank the support of the companies involved in the development phase, namely Full Body Insight, Oncovision and Quibim. Authors also thank the engineering team at La Fe hospital during the installation and the patient's recruitment team for their support and kindness. Moreover, authors also would like to acknowledge the help of the PETsys members in the implementation of the online energy threshold.

## REFERENCES

[1] S. Vandenberghe, et al., "State of the art in total body PET," EJNMMI physics 20, 35, 2020.

[2] Lecoq P., "Pushing the limits of time-of-flight PET imaging," IEEE Trans Radiat Plasma Med Sci. 2017;1(6):473–485

[3] S.R. Cherry, et al., "Total-Body PET: Maximizing Sensitivity to Create New Opportunities for Clinical Research and Patient Care," J. Nucl. Med. 59, 3-12, 2018.

[4] S. Surti, et al., "Total Body PET: Why, How, What for?," IEEE Trans. Rad. Plasma Med. Scie, 4, 283-292, 2020.

[5] X. Sun, at al., "11C-CFT PET brain imaging in Parkinson's disease using a total-body PET/CT scanner," EJNMMI Physics 11, 40, 2024.

[6] E.C. Dijkers, et al., "Biodistribution of $^{89}$Zr-trastuzumab and PET Imaging of HER2-Positive Lesions in Patients With Metastatic Breast Cancer," Clin. Pharmacol. Ther. 87, 586–592, 2010.

[7] G.A. Prenosil, et al., "Performance Characteristics of the Biograph Vision Quadra PET/CT System with a Long Axial Field of View Using the NEMA NU 2-2018 Standard," J. Nucl. Med. 63, 476 - 484, 2022.

[8] H. Zhang, et al., "Performance Characteristics of a New Generation 148-cm Axial Field-of-View uMI Panorama GS PET/CT System with Extended NEMA NU 2-2018 and EARL Standards," J. Nucl. Med. 65, 1974 - 1982, 2024.

[9] R.D. Badawi, et al., "First Human Imaging Studies with the EXPLORER Total-Body PET Scanner," J. Nucl. Med. 60, 299 – 303, 2019.

[10] B. Dai, et al., "Performance evaluation of the PennPET explorer with expanded axial coverage," Phys. Med. Biol. 68, 095007, 2023.

[11] P. Moskal, et al., "Simulating NEMA characteristics of the modular total-body J-PET scanner—an economic total-body PET from plastic scintillators," Phys. Med. Biol. 66, 175015, 2021.

[12] P. Moskal, et al., "Positronium image of the human brain in vivo" Science Advances 10, eadp2840, 2024.

[13] R.S. Miyaoka and A.L. Lehnert, "Small animal PET: a review of what we have done and where we are going," Phys. Med. Biol. 65, 24TR04, 2020.

[14] S. Surti and J.S. Karp, "Impact of detector design on imaging performance of a long axial field-of-view, whole-body PET scanner," Phys. Med. Biol. 60, 5343–5358, 2015.

[15] J.P. Schmall, J.S. Karp, M. Werner and S. Surti, "Parallax error in long-axial field-of-view PET scanners—a simulation study," Phys. Med. Biol. 61, 5443–5455, 2016.

[16] NEMA 2018 NU 2-2018-performance measurements of positron emission tomographs National Electrical Manufacturers Association. Rosslyn, USA.

[17] N. Cucarella, et al., "Timing evaluation of a PET detector block based on semi-monolithic LYSO crystals," Med. Phys. 48, 8010–8023, 2021.

[18] F. Muller, at al., "A semi-monolithic detector providing intrinsic DOI-encoding and sub-200 ps CRT TOF-capabilities for clinical PET applications," Med. Phys. 49, 7469-7488, 2022.

[19] J. Barrio, et al., "Time and energy characterization of semi-monolithic detectors with different treatments and SiPMs suitable for clinical imaging," IEEE Trans. Rad. Plasma Med. Scie. 7, 785, 2023.

[20] Y. Zhao, et al., "Evaluation of a new type of semi-monolithic DOI detector with different surface treatments." IEEE Nucl. Scie. Sym. Med. Imag Conf (NSS/MIC), 2023.

[21] G. Cañizares, et al., "Simulation study of clinical PET scanners with different geometries, including TOF and DOI capabilities," IEEE Trans. Rad. Plasma Med. Scie. 8.6: 690-699, 2024.

[22] M. Freire, et al., "Position estimation using neural networks in semi-monolithic PET detectors," Phys. Med. Biol. 67.24, 255011, 2022.

[23] D. Sanchez, A. Gonzalez-Montoro, J. Barbera, et al., "Design and Evaluation of a High-Performance Readout for Total-Body PET," IEEE Nucl. Scie. Sym. Med. Imag Conf (NSS/MIC), 2022.

[24] F. Pagano, et al. "Semi-Monolithic Detectors for TOF-DOI Brain PET: Optimization of Time, Energy, and Positioning Resolutions With Varying Surface Treatments," IEEE Trans. Rad. Plasma Med. Scie., accepted 2025.

[25] A. Gonzalez-Montoro, et al., "First Results of the 4D-PET Brain System," IEEE Trans. Rad. Plasma Med. Scie. 8, 839, 2024.

[26] R. Bugalho et al., "Experimental characterization of the TOFPET2 ASIC," J. Instrum. 15, P03029, 2019.

[27] D. Sanchez, et al., "IMAS: a full-body PET system with enhanced TOF and DOI capabilities," IEEE Nucl. Scie. Sym. Med. Imag Conf (NSS/MIC), 2024.

[28] A.E. Perkins, et al. "Time of flight coincidence timing calibration techniques using radioactive sources," IEEE Nucl. Scie. Sym. Med. Imag Conf (NSS/MIC), 2005.

[29] T. Yamaya et al., "DOI-PET image reconstruction with accurate system modeling that reduces redundancy of the imaging system," IEEE Trans. Nucl. Sci., 50, 1404 – 1409, 2003.

[30] S. Vandenberghe, et al., "Recent developments in time-of-flight PET," EJNMMI Phys. 3, 1 – 30, 2016.

[31] M. Filipović, et al., "Time-of-flight (TOF) implementation for PET reconstruction in practice," Phys. Med. Biol. 64, 23NT01, 2019.

[32] M. Conti et al., "First experimental results of time-of-flight reconstruction on an LSO PET scanner," Phys. Med. Biol. 50, 4507 – 4526, 2005.

[33] M.E. Daube-Witherspoon, V. Viswanath, M.E. Werner, J.S. Karp, "Performance Characteristics of Long Axial Field-of-View PET Scanners with Axial Gaps," IEEE Trans. Rad. Plasma Med. Sci. 5, 322 – 330, 2021.

[34] S. Surti, et al., "Performance of Philips Gemini TF PET/CT Scanner with Special Consideration for Its Time-of-Flight Imaging Capabilities," J. Nucl. Med. 48, 471 – 480, 2007.

[35] S. Klein, M. Staring, K. Murphy, M.A. Viergever, J.P.W. Pluim, "elastix: A Toolbox for Intensity-Based Medical Image Registration," IEEE Trans. Med. Imaging 29, 196 – 205, 2010.

[36] B.A. Spencer, et al., "Performance Evaluation of the uEXPLORER Total-Body PET/CT Scanner Based on NEMA NU 2-2018 with Additional Tests to Characterize PET Scanners with a Long Axial Field of View," J. Nucl. Med. 62, 861 – 870, 2021.